\documentclass[journal]{IEEEtran}

\usepackage{graphicx}

\usepackage{cite}
\usepackage{amsmath}

\usepackage[section]{placeins}
\usepackage{ textcomp }
\usepackage{xcolor}

\begin{document}

\title{Dual-wavelength lasers on generic foundry platform}

\author{Robert~Pawlus,~Robbe~de~Mey,~Stefan~Breuer,~and~Martin~Virte,~\IEEEmembership{Member,~IEEE,}
\thanks{R. Pawlus, R. de Mey and M. Virte are with the Brussels Photonics Team, Department of Applied Physics and Photonics (TONA/B-PHOT), Vrije Universiteit Brussel, 1050 Brussels, Belgium (rpawlus@b-phot.org, mvirte@b-phot.org)}
\thanks{R. Pawlus and S. Breuer is with the Institute of Applied Physics, Technische Universität Darmstadt, Schlossgartenstraße 7, 64289 Darmstadt, Germany (stefan.breuer@physik.tu-darmstadt.de)}
\thanks{The authors acknowledge support from the Research Foundation - Flanders (FWO) (grants 1530318N, G0G0319N), and the METHUSALEM program of the Flemish Government, the Adolf Messer Foundation (personal grant) and the German Research Foundation (DFG) (389193326). We acknowledge support by the German Research Foundation and the Open Access Publishing Fund of Technische Universit\"{a}t Darmstadt.}}

\maketitle

\begin{abstract}

We propose and implement four simple and compact dual-wavelength laser concepts integrated in a Photonic Integrated Circuit (PIC) based on a InP generic foundry platform.  In a first step, we arrange two detuned Distributed-Bragg-Reflectors (DBR) in either a sequential or in a parallel order, acting as narrowband wavelength selective cavity mirrors. In a second step, we close the cavities by using either a third DBR or by using a Multimode-Interference-Reflector (MIR). We present LI-characteristics and optical spectra emitting around 1550~nm with wavelength separations of 1~nm or 10~nm and evaluate their particular potential for simultaneous dual-wavelength emission. In addition, we find either one or multiple equal power points as well as complete switches when the gain current is being tuned. We discuss the characteristics and limitations of each concept including arranging the detuned DBRs in a sequential or parallel order.
\end{abstract}


\section{Introduction}

\IEEEPARstart{D}{ual} wavelength semiconductor lasers gained interest in recent years for various applications related to THz and mmWave generation \cite{Hoffmann2005} \cite{Huang2009}, velocimetry \cite{Gioannini2014a} or all optical signal processing \cite{Wu2016}. Different techniques have been developed to generate dual-wavelength emission. The simplest way is to combine the beams of two separate semiconductor laser sources, which usually results in bulky setups and poses a challenge to stabilize mechanically. Although this allows for a great flexibility in spectral tuning and the precise control of their optical power, semiconductor lasers with intrinsic multi-wavelength selection like edge-emitting quantum-dot lasers can emit from the ground and excited state at the same time, yet usually require asymmetrical biasing \cite{Markus2006}, low temperatures \cite{Pawlus2017} or embedded gratings \cite{Naderi2010a} to achieve a dual-wavelength emission. External optical grating feedback adds an additional degree of emission control to extend their operation range \cite{Virte2014} \cite{Virte2016}, but demands high mechanical and thermal stability. In \cite{Huang2004}, dual wavelength emission in the 1300-1500~nm region, separated by up to 174~nm, was demonstrated using an external cavity, however, with a challenging alignment for the individual modes. Lasers with integrated optical feedback promise a higher thermal and mechanical stability. A dual-wavelength DBR laser emitting around 1080~nm was demonstrated with a tuneable dual-wavelength emission between 0.3~nm and 6.9~nm in \cite{Roh2000}.
Generic foundry platforms offer an easy and affordable access to a mass manufacturing process and customized laser sources can be designed.
Different multi-wavelength semiconductor lasers emitting around 1550~nm have already been demonstrated: In \cite{Carpintero2012}, multiple lasers were combined to achieve a 16-channel laser source with spectral separations of 0.8~nm. In \cite{Docter2010}, a four-channel Fabry-Perot laser spectrally deparated by 3.2~nm and in \cite{Ermakov2012}, a four-channel ring laser with 1.336~nm spectral separation has been demonstrated, both based on integrated external feedback as the control mechanism. Although these lasers are already integrated and can emit on multiple wavelengths, they are based on arrayed-waveguide-gratings as the wavelength selective element which poses challenges in the design processes as they have a large foot print and lack in ease of use due to multiple gain sections and control currents.

In this work, we propose four simple and compact dual-wavelength lasers concepts and demonstrate their implementation onto the generic multi project wafer platform offered by SmartPhotonics. Using their provided set of building blocks, our designs are based on two different wavelength separations of either 1~nm or 10~nm. The gain-current serves as the only parameter to control their emission properties. We expect the lasers to exhibit a controllable switch from one longitudinal mode to the other due to the blue shift of the gain-spectrum as the gain current is being tuned. Our goal is to achieve a simultaneous dual-wavelength emission at these transition points. Although we obtain dual-wavelength emission for all designs, we clearly observe that simultaneous emission and switching behavior vary particularly from design to design.

\section{Dual-wavelengths laser designs}

\begin{figure}[t]
\centering
\includegraphics[width=3.4in]{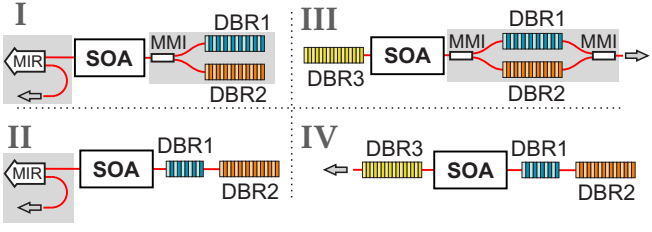}
\caption{Dual wavelength laser layouts: I) two parallel DBRs, connected via an MMI to an SOA, form a cavity with a MIR, II) two sequential DBRs forming a cavity with a MIR, III) two parallel DBRs forming a cavity with another DBR and IV) two sequential DBRs forming a cavity with another DBR. Gray area indicate deeply etched components. Not shown are transition sections from shallow/deep etching, isolation sections and mode filters.}
\label{FIG: Laser Layouts}
\end{figure}

The four layouts implemented and studied in this work are depicted in Fig. \ref{FIG: Laser Layouts}. Two narrowband DBRs (DBR$_1$ \& DBR$_2$) act as the wavelength selective elements and are arranged either in parallel (I \& III) or sequentially (II \& IV). To close the cavity, either a broadband MIR (I \& II), or a third DBR (DBR$_3$ in III \& IV) is used. The layouts based on a MIR allow for a large and variable spectral wavelength separation within the gain bandwidth. A precise design for the DBR reflectivities is essential to achieve an equal gain for a particular wavelength separation. The DBRs of our MIR based layouts are designed to have a spectral separation of 10~nm to avoid any interference between the DBRs. On the other hand, the layouts based on three DBRs use a much wider bandwidth for DBR$_3$ which reflects at both wavelengths generated by DBR$_1$ and DBR$_2$. The required spectral overlap limits the wavelength splitting to about 1~nm but allows for a loss control when their wavelengths and subsequently their spectral overlap is tuned.
\begin{table}[t]
\centering
\caption{Implemented DBR parameter for different laser layouts.}
\begin{tabular}{c|c|c|c|c|c}
Layout & \begin{tabular}{@{}c@{}} $\Delta \lambda$ \\ (nm) \end{tabular} & \begin{tabular}{@{}c@{}} Design \\ wavelength (nm)\end{tabular} & \begin{tabular}{@{}c@{}} DBR length \\ ($\mu$ m)\end{tabular} & R & \begin{tabular}{@{}c@{}} BDW \\ (nm) \end{tabular} \\ \cline{1-6}
 
 I & 10 & \begin{tabular}{@{}c@{}} $\lambda_1=1535.0$ \\ $\lambda_2=1545.0$  \end{tabular}  & \begin{tabular}{@{}c@{}} $DBR_1=500$ \\ $DBR_2=500$ \end{tabular}  &  \begin{tabular}{@{}c@{}} 0.97 \\ 0.97 \end{tabular}   &  \begin{tabular}{@{}c@{}} 1.08 \\ 1.08 \end{tabular} \\ \cline{1-6}

 II & 10 & \begin{tabular}{@{}c@{}} $\lambda_1=1545.0$ \\ $\lambda_2=1535.0$  \end{tabular} & \begin{tabular}{@{}c@{}} $DBR_1=250$ \\ $DBR_2=500$ \end{tabular}  &  \begin{tabular}{@{}c@{}}  0.72 \\ 0.97 \end{tabular}  &  \begin{tabular}{@{}c@{}}  1.47 \\ 1.08 \end{tabular} \\  \cline{1-6}

 III & 1 & \begin{tabular}{@{}c@{}} $\lambda_1=1541.3$ \\ $\lambda_2=1539.3$ \\ $\lambda_3=1540.0$ \end{tabular} & \begin{tabular}{@{}c@{}} $DBR_1=450$ \\ $DBR_2=450$ \\ $DBR_3=480$ \end{tabular} &  \begin{tabular}{@{}c@{}} 0.96 \\ 0.96 \\ 0.97 \end{tabular} &  \begin{tabular}{@{}c@{}} 1.11 \\ 1.11 \\ 1.09 \end{tabular} \\ \cline{1-6}

 IV & 1 & \begin{tabular}{@{}c@{}} $\lambda_1=1541.3$ \\ $\lambda_2=1539.3$ \\ $\lambda_3=1540.0$ \end{tabular} & \begin{tabular}{@{}c@{}} $DBR_1=200$ \\ $DBR_2=350$ \\ $DBR_3=250$ \end{tabular}  &  \begin{tabular}{@{}c@{}} 0.58 \\ 0.89 \\ 0.72 \end{tabular} &  \begin{tabular}{@{}c@{}} 1.72 \\ 1.23 \\ 1.47 \end{tabular}
\label{TAB: Table with DBR variables}
\end{tabular}
\end{table}
In layouts I \& II, 2-port MIRs are used to provide a reflectivity of about 0.4 for the laser cavities. The parameters for all DBRs were determined using the PIC simulator Lumerical Interconnect, where each layout was optimized for dual-wavelength emission with equal optical powers. The design wavelengths $\lambda_1$, $\lambda_2$ \& $\lambda_3$ for each DBR are listed in Tab. \ref{TAB: Table with DBR variables}, together with their lengths, reflectivities and bandwidths, determined for each laser cavity. DBR$_1$ in layouts II \& IV were implemented with a short length to achieve a large free spectral range for both cavities - created by DBR$_1$ \& DBR$_2$ - to be single mode. The coupling coefficient is fixed to 50~$cm^{-1}$ and the maximal DBR length to 500~$\mu$m. For layouts III \& IV, we implemented a 0.3~nm longer wavelength for DBR$_1$ \& DBR$_2$ onto the PIC compared to the wavelengths of $\lambda_1=1539$~nm and $\lambda_2=1540$~nm used in the simulations as their wavelengths can be tuned experimentally. As DBR$_3$ in layout III has a high reflectivity and DBR$_1$ \& DBR$_2$ are transparent to each other, we combined their outputs with a multi-mode interference (MMI) coupler to maximize the optical output power. To achieve the most compact footprint on the PIC, we used deeply etched components wherever beneficial, shown by a gray background in Fig. \ref{FIG: Laser Layouts}. All SOAs have a length of 500~$\mu$m and are operated at a maximal gain current of 90~mA, however, we limit ourselves to 80-85~mA to avoid excessive load. Intracavity modefilters have been implemented for layouts II, III and IV, for layout I the modefilter had to be discarded to reduce the laser cavity length to achieve a single-mode emission.

\section{Experimental results}

A Thorlabs Pro8 system controls the current and temperature of the gain section and the PIC. An Apex AP2083A (resolution down to 5~MHz~/~40~fm) allows for measuring the optical spectra as well as the optical power. All measurements were performed at a temperature of 20\textdegree C. A lensed fiber was used to couple the light out of the PIC. The presented results were performed on three different PICs from three different wafers which provides also insight into the robustness and reproducibility of each layout. Key results for each layout are given in Table \ref{TAB: Table with exp. results} with the spectral separation $\Delta \lambda$ and the wavelengths acquired experimentally together with the best values found for the optical power, the side mode suppression ratio (SMSR) and the longitudinal mode separation.

\begin{figure}[h]
\includegraphics[height=0.455\columnwidth]{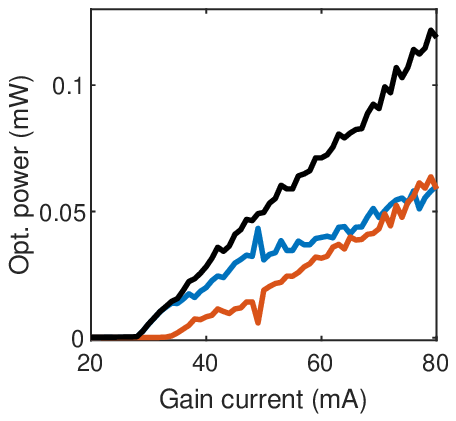}
  \put(-55,104){(a)}
\centering
  \hfill\includegraphics[height=0.463\columnwidth]{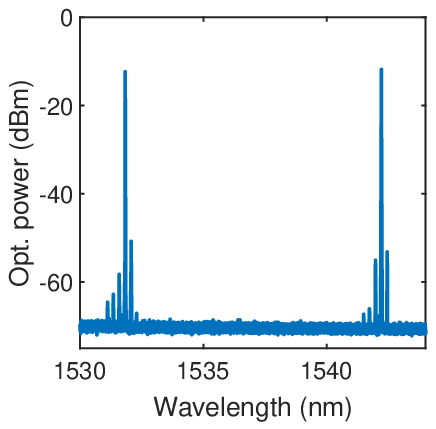}
    \put(-55,104){(b)}

    \caption{Layout I: Wavelength resolved LI curve (a), shown are the shorter and longer wavelengths in blue and red, respectively, the total power is shown in black. The optical spectrum (b) is shown at a gain current of 80~mA.}
    \label{FIG: I}
\end{figure}

\FloatBarrier
\textbf{Layout I:} The measured LI curve is depicted in Fig.~\ref{FIG: I}~(a) without current applied to any DBR.
Lasing starts at 28~mA at the shorter wavelength of 1531.8~nm, corresponding to DBR$_1$. At 35~mA, the longer wavelength of 1542.2~nm starts to emit, corresponding to DBR$_2$. Both modes behave similarly with an increase in optical power and show a wide range of equal optical power between 65 and 80~mA. At a gain current of 49~mA, a switch to another longitudinal mode occurs within the longer wavelengths. The highest optical power, with about \mbox{-12~dBm} for each mode, is achieved at a gain current of 80~mA, the corresponding optical spectrum is depicted in Fig.~\ref{FIG: I}~(b). The separation of the two modes is 10.4~nm with multiple side modes appearing for each wavelength with a SMSR of at least 38.5~dB. Although this laser seems like the ideal choice for a dual-wavelength laser, across different PICs only a single mode emission occured, suggesting a high requirement for smaller DBR tolerances for this layout.
 
\FloatBarrier
\FloatBarrier

\textbf{Layout II:} The currents were set to 5~mA for DBR$_1$ and 4~mA for DBR$_2$ to achieve a single mode emission for each  
\begin{figure}[h]
    \centering
  \centering
  \includegraphics[height=0.46\columnwidth]{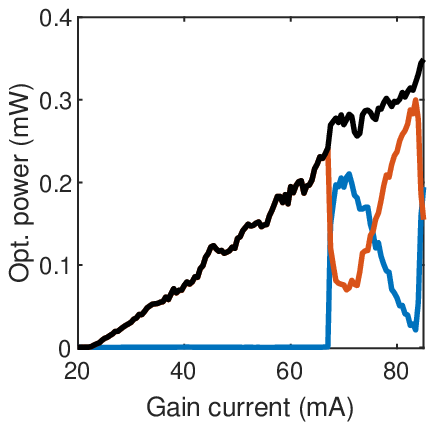}
    \put(-55,104){(a)}
  \centering
  \hfill\includegraphics[height=0.46\columnwidth]{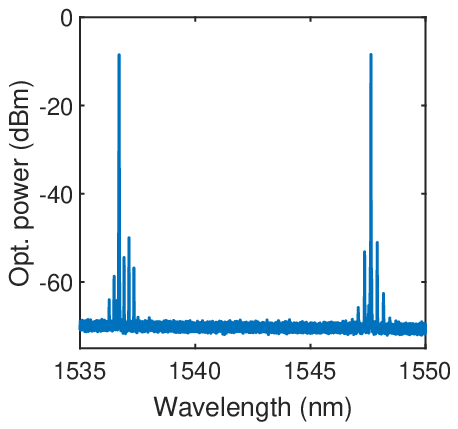}
    \put(-55,104){(b)}
    \caption{Layout II: Wavelength resolved LI curve (a), shown are the shorter and longer wavelengths in blue and red, respectively, the total power is shown in black. The optical spectrum (b) is shown at a gain current of 85~mA.}
    \label{FIG: II}
\end{figure}
wavelength. The resulting LI curve is depicted in Fig.~\ref{FIG: II}~(a). The laser starts to emit from the longer wavelength of 1547.6~nm and at a gain current of 67~mA, the shorter wavelength of 1536.7~nm starts to emit and increases rapidly in optical power while the longer wavelength experiences a drop. With a further increase of the gain current, the two modes compete with each other, resulting in multiple power exchanges. The optical spectrum at a gain current of 85~mA is depicted in Fig.~\ref{FIG: II}~(b) with a wavelength separation of 10.9~nm, an equal optical output power of about -8.5~dBm and a SMSR of at least 41.5~dB. The wavelengths are shifted 1.5~nm to longer wavelengths, indicating high variances in the target wavelengths of the DBRs across the wafers. Multiple equal power points are found by this layout, changing the DBR currents does not change the overall behavior of the multiple power exchanges, but shifts the equal power points slightly to different currents. This layout is best suited to achieve a simultaneous dual-wavelength emission as reproducible behavior across the different chips could be found.
\FloatBarrier
\FloatBarrier
\textbf{Layout III:} Due to the implementation of longer wavelengths for DBR$_1$ and DBR$_2$ of 0.3~nm mentioned above, a current of 0~mA, 2~mA and 8~mA had to be injected to DBR$_1$, DBR$_2$ and DBR$_3$, respectively, to achieve the LI 
\begin{figure}[h]
    \centering
  \centering
  \includegraphics[height=0.46\columnwidth]{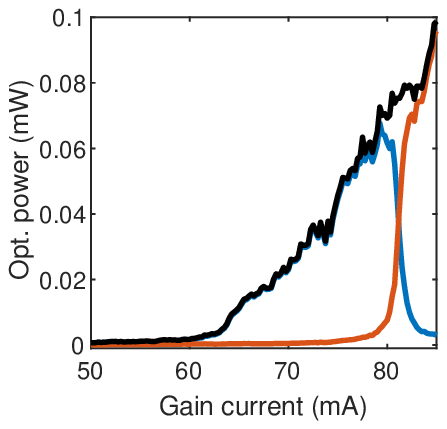}
    \put(-55,104){(a)}
  \centering
  \hfill\includegraphics[height=0.46\columnwidth]{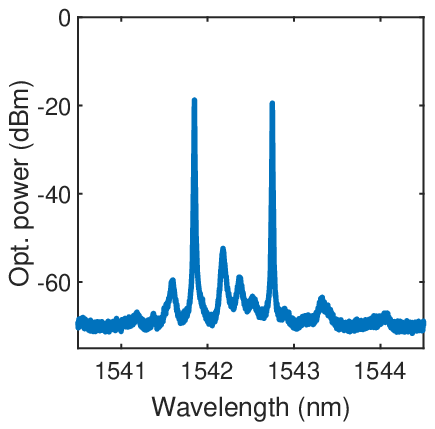}
    \put(-55,104){(b)}
    \caption{Layout III: Wavelength resolved LI curve (a), shown are the shorter and longer wavelengths in blue and red, respectively, the total power is shown in black. The optical spectrum (b) is shown at a gain current of 81~mA.}
    \label{FIG: III}
\end{figure}
curve depicted in Fig.~\ref{FIG: III}~(a). This laser has a high lasing threshold, starting to emit on the shorter wavelength of 1541.8~nm. At a gain current of 80~mA, the longer wavelength of 1542.7~nm starts to emit and reaches an equal power point at 81~mA. Beyond this point, the laser is only emitting on the longer wavelength with a steeper slope in the optical power up to 85~mA. The optical spectrum at the equal power point is depicted in Fig. \ref{FIG: III} (b) and shows the equal power point of -19~dBm for each of the modes with a SMSR of at least 32.8~dB. This layout turned out to be the most versatile as a simultaneous dual wavelength emission as well as a full wavelength switch could be achieved across different PICs.

\FloatBarrier

\begin{figure}[h]
    \centering
  \centering
  \includegraphics[height=0.455\columnwidth]{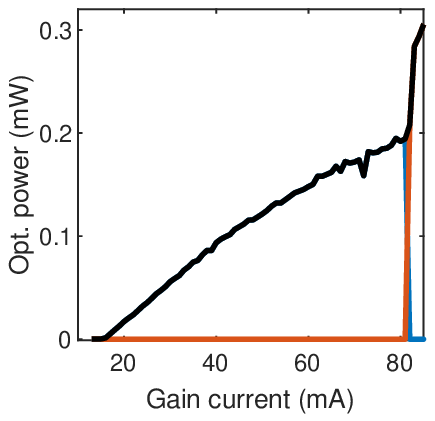}
    \put(-55,104){(a)}
  \centering
  \hfill\includegraphics[height=0.465\columnwidth]{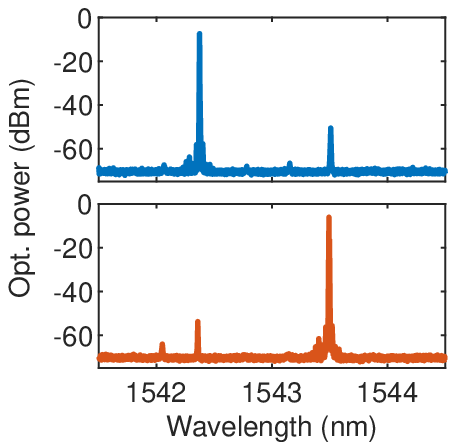}
    \put(-54,104){(b)}
    \put(-54,55){(c)}
    \caption{Layout IV: Wavelength resolved LI curve (a), shown are the shorter and longer wavelengths in blue and red, respectively, the total power is shown in black. The optical spectrum is shown at a gain current of 81~mA (b, top) and 82~mA (b, bottom).}
    \label{FIG: IV}
\end{figure}

\textbf{Layout IV:} We only injected a current of 1.7~mA to DBR$_3$ to achieve the LI curve depicted in Fig. \ref{FIG: IV} (a).
This laser has a low threshold and starts to emit on the shorter wavelength of 1542.4~nm. With increasing gain current, the optical power increases while showing indications of a gain saturation with a flattening of the LI curve. At a gain current of 82~mA, a sudden switch to the longer wavelength of 1543.5~nm occurs, exhibiting also a hysteresis cycle. The optical spectra for a gain current of 81~mA and 82~mA are depicted in Fig. \ref{FIG: IV} (b) and (c), respectively. The optical power at the switching point is -7~dBm in single-mode emission with the other mode suppressed by at least 43.2~dB. In this layout  we achieve a direct switch from one wavelength to the other and find the same behavior in all of our PICs, making this laser an attractive choice for dual-wavelength switching.

\section{Discussion}
All proposed structures except layout II start to emit on the shorter wavelength, followed by emission on the longer wavelength. This can be altered by the DBR reflectivities, the applied DBR currents (layout III \& IV) or the influence of a temperature gradient caused by biasing a neighboring SOA. The change in emission from the longer to the shorter wavelength in layout II can be explained by the blue shift of the gain spectrum with increasing gain current. The tuning of the gain current induces a shift of the whole spectrum by about 0.04~nm/mA to longer wavelengths due to the change in refractive index by the higher current density.
All structures showed simultaneous or sequential dual-wavelength emission even though reproducibility of layout I is poor. Each of the layouts II, III \& IV showed reproducible results across different PICs and wafers and suggest a good robustness of the designs.
Layouts I \& III with a parallel DBR arrangement are simple to design due to their symmetric composition and can have the same large DBR lengths with high reflectivities. However, their transparency to each other results in high intracavity losses which have to be compensated by the SOA and therefore result in lower optical output powers. Layout I seems to be prone to deviations in the DBR reflectivities which can lead to a different gain for each wavelength and a potential single mode emission. This asymmetric gain can be compensated in layout III by the tuning of the DBRs. 
The layouts II \& IV with sequentially arranged DBRs show lower losses but require a careful design in their lengths to provide equal gain for both wavelengths. They showed to have the highest optical output power at multiple equal optical power points for layout II and at a single mode emission for layout IV.
For the MIR based layouts I \& II, single mode emission for both wavelengths is achieved, confirming this layout to be a versatile approach for a varying wavelength separations.
In the DBR based layouts III \& IV, active control of each wavelength is achieved by the relative tuning of the DBRs. Depending on the set of currents for the DBRs, a single mode emission for both wavelengths as well as the desired dual wavelength emission is possible. 
The combined outputs of DBR$_1$ \& DBR$_2$ in layout III were implemented to study their influence on the laser output. As the laser has a low output power, using this approach could be beneficial to improve the performance of layout I. Using a 1-port MIR to close the cavity could result in lower losses and a higher optical output power.
The temperature of the PIC has a major impact on the DBR performance, wavelength shifts ranging from 0.11~nm/\textdegree C to 0.19~nm/\textdegree C to longer wavelengths when increasing the temperature could be found. Hence, controlling the temperature can be used to match the experimental to the expected wavelengths. An extensive study on the DBR tuning for each layout is however left for further investigation.
\section{Conclusion}
\begin{table}[ht]
\centering
\caption{Key experimental results for each Layout.}
\begin{tabular}{c|c|c|c|c|c}

Layout & 
\begin{tabular}{@{}c@{}} $\Delta \lambda$ \\ (nm) \end{tabular} &
\begin{tabular}{@{}c@{}} Experimental\\ wavelength \\ (nm)\end{tabular} &
\begin{tabular}{@{}c@{}} Fiber coupled \\ optical power \\ (dBm) \end{tabular} &
\begin{tabular}{@{}c@{}} SMSR \\ (dB) \end{tabular} &
\begin{tabular}{@{}c@{}} long. mode \\ separation \\ (nm) \end{tabular} \\ \cline{1-6}

I &
10.4 & 
\begin{tabular}{@{}c@{}} 1531.8 \\ 1542.2 \end{tabular} &
\begin{tabular}{@{}c@{}} -12.3 \\ -11.8 \end{tabular} &
\begin{tabular}{@{}c@{}} 38.5 \\ 41.4 \end{tabular} &
\begin{tabular}{@{}c@{}} 0.24 \\ 0.24 \end{tabular} \\ \cline{1-6}

II & 
10.9 & 
\begin{tabular}{@{}c@{}} 1536.7 \\ 1547.6 \end{tabular} &
\begin{tabular}{@{}c@{}} -8.5 \\ -8.4 \end{tabular} &
\begin{tabular}{@{}c@{}} 41.5 \\ 42.7 \end{tabular} &
\begin{tabular}{@{}c@{}} 0.27 \\ 0.21 \end{tabular} \\ \cline{1-6}

III & 
0.9 &
\begin{tabular}{@{}c@{}} 1541.8 \\ 1542.7 \end{tabular} &
\begin{tabular}{@{}c@{}} -18.8 \\ -19.4 \end{tabular} &
\begin{tabular}{@{}c@{}} 33.4 \\ 32.8 \end{tabular} &
\begin{tabular}{@{}c@{}} 0.25 \\ 0.25 \end{tabular} \\ \cline{1-6}

IV & 
1.1 &
\begin{tabular}{@{}c@{}} 1542.4 \\ 1543.5 \end{tabular} &
\begin{tabular}{@{}c@{}} -7.4* \\ -6.1* \end{tabular} &
\begin{tabular}{@{}c@{}} 43.2 \\ 47.8 \end{tabular} &
\begin{tabular}{@{}c@{}} 0.37 \\ 0.31 \end{tabular}

\label{TAB: Table with exp. results}
\end{tabular}
           *Optical power for single-mode emission
\end{table}
To conclude, we presented four different compact laser concepts capable of achieving simultaneous dual-wavelength emission at 1550~nm. Key results are summarized in Tab. \ref{TAB: Table with exp. results}. We exploited the gain current as the dominant control parameter and were able to achieve one or multiple equal power points as well as a full switch between the two wavelengths. We highlighted the unique characteristics of each approach to form the laser cavities, either offering a wide spectral wavelength separation or an active control over each wavelength. The sequential MIR layout II allowed for multiple equal power points with the highest optical output power and is our recommended choice for an individual wavelength separation within the gain bandwidth. For wavelength separations around 1~nm, layout III turned out to be the most versatile solution, allowing for simultaneous dual-wavelength emission and switching while layout IV is the most suitable for exclusive switching applications with the highest optical power in single-mode emission. 

\bibliographystyle{IEEEtran}

\begin{IEEEbiographynophoto}{Robert Pawlus}
received his Master of Science degree in Physics from the Technische Universität Darmstadt, Germany in 2016 and is currently pursuing his PhD degree at the Brussels Photonics Team (B-PHOT) of the Vrije Universiteit Brussel, Belgium.\\
\\
\textbf{Robbe de Mey}
received his Master degree in Photonics from the Brussels Photonics Team (B-PHOT) of the Vrije Universiteit Brussel, Belgium in 2019 and is currently pursuing his PhD degree at the same university.\\
\\
\textbf{Stefan Breuer}
received the Ph.D. degree in physics from Technische Universität Darmstadt, Darmstadt, Germany, in 2010. He is currently a Post-Doctoral Research Fellow with Technische Universität Darmstadt.\\
\\
\textbf{Martin Virte}
received his master of Engineering from the French “Grande Ecole” Supélec (now CentraleSupélec) in 2011. In 2014, he received his PhD in engineering  from both VUB and Supélec as part of a joint PhD program. From 2014 to 2018, he was a post-doctoral researcher at VUB. Since 2018, he is a research professor with the Brussels Photonics Team (B-PHOT) of the Vrije Universiteit Brussel, Belgium. 
\end{IEEEbiographynophoto}

\end{document}